\documentclass[12pt,preprint]{aastex}

\newcommand{\kms}{km\,s$^{-1}$}

\newcommand{\Ha}{H$\alpha$}

\newcommand{\CII}{C\,{\sc ii}} 
\newcommand{\CI}{C\,{\sc i}}

\newcommand{\SiII}{Si\,{\sc ii}}
\newcommand{\SiI}{Si\,{\sc i}}
\newcommand{\HeI}{He\,{\sc i}}
\newcommand{\FeII}{Fe\,{\sc ii}}
\newcommand{\MgII}{Mg\,{\sc ii}}
\newcommand{\NaID}{Na\,{\sc i}~D}
\newcommand{\NeI}{Ne\,{\sc i}}

\newcommand{\Mo}{M$_{\odot}$}

\shorttitle{The extreme case of SN\,2007gr}
\shortauthors{Valenti et al.}

\begin{document}
\setcounter{footnote}{-1}

\title{The carbon-rich type Ic SN\,2007gr: the photospheric phase}


\author{S.   Valenti\altaffilmark{1,2}; N. Elias-Rosa\altaffilmark{3};
 S.   Taubenberger\altaffilmark{3};  V.   Stanishev\altaffilmark{4}; 
 I.   Agnoletto\altaffilmark{5};    D. Sauer\altaffilmark{3}; 
E. Cappellaro\altaffilmark{5};  A.   Pastorello\altaffilmark{6};  
S.  Benetti\altaffilmark{5};  A. Riffeser \altaffilmark{7,8}; 
 U. Hopp \altaffilmark{7,8};  H. Navasardyan\altaffilmark{5}; 
 D. Tsvetkov\altaffilmark{9};  V. Lorenzi\altaffilmark{10};
F. Patat\altaffilmark{2}; 
 M. Turatto\altaffilmark{5}; R. Barbon \altaffilmark{11};
S. Ciroi \altaffilmark{11};  F. Di Mille\altaffilmark{11}
S. Frandsen \altaffilmark{12}; J.P.U. Fynbo \altaffilmark{13};
P. Laursen \altaffilmark{13}; P.A. Mazzali\altaffilmark{3,14}}

\altaffiltext{1}{Dipartimento di Fisica, Universit\'a di Ferrara, 
via Saragat 1, 44100 Ferrara, Italy}
\altaffiltext{2}{European Organisation for Astronomical Research in the
Southern Hemisphere, Karl-Schwarzschild-Str. 2, 85748 Garching 
bei M\"unchen, Germany; svalenti@eso.org}
\altaffiltext{3}{Max-Planck-Institut f\"ur Astrophysik,
Karl-Schwarzschild-Str. 1, 85741 Garching bei M\"unchen, Germany}
\altaffiltext{4}{Department of Physics, Stockholm University, 
Stockholm, Sweden}
\altaffiltext{5}{INAF-Osservatorio Astronomico, vicolo
dell'Osservatorio 5, 35122 Padova, Italy}
\altaffiltext{6}{Astrophysics Research Centre, School of Mathematics and
Physics, Queen's University Belfast, Belfast BT7 1NN, UK}
\altaffiltext{7}{Universit\"ats-Sternwarte M\"unchen, Scheinenstr. 1 
81679  M\"unchen, Germany}
\altaffiltext{8}{Max-Planck-Institut f\"ur Extraterrestrische Physik,
Giessenbachstr, 85748 Garching bei M\"unchen, Germany}
\altaffiltext{9}{Sternberg Astronomical Institute, Moscow University,
Univeritetskii pr.13, Moscow 119992, Russia}
\altaffiltext{10}{Fundaci\'on Galileo Galilei-INAF, Telescopio Nazionale Galileo, E-38700 Santa Cruz de la Palma, Tenerife, Spain }
\altaffiltext{11}{Dipartimento di Astronomia, Universit\'a di Padova, vicolo dell'Osservatorio 2, 35122 Padova, Italy}
\altaffiltext{12}{Institut for Fysik og Astronomi, Aarhus Universitet, Ny 
Munkegade, Bygn. 1520, 8000 Aarhus C}
\altaffiltext{13}{Dark Cosmology Centre, Niels Bohr Institute, University of Copenhagen, Juliane Maries Vej 30, 2100 Copenhagen, Denmark}
\altaffiltext{14}{INAF - Osservatorio Astronomico, Via Tiepolo 11, 34143 Trieste, Italy}


\begin{abstract}
The first  two months of  spectroscopic and photometric  monitoring of
the nearby type Ic SN\,2007gr are presented.  The very early discovery
(less  than 5  days  after  the explosion)  and  the relatively  short
distance  of  the host  galaxy  motivated  an extensive  observational
campaign.  SN\,2007gr  shows an average peak  luminosity but unusually
narrow  spectral  lines  and  an  almost  flat  photospheric  velocity
profile.  The detection of prominent carbon features in the spectra is
shown   and   suggest   a   wide   range  in   carbon   abundance   in
stripped-envelope supernovae.  SN 2007gr  may be an important piece in
the puzzle of the observed diversity of CC SNe.
\end{abstract}



\section{Introduction}\label{intro}

Massive  stars  (M  $>$  8  \Mo)  end  their  lives  as  core-collapse
supernovae (CC SNe).   SNe resulting from stars that  had lost their H
envelopes  before they  collapse are  called stripped-envelope  CC SNe
\citep[cf.][]{clocchiatti96,filippenko97}    and   are   traditionally
classified as type Ib if the He envelope was still present and type Ic
if it had been lost.

Stripped-envelope CC-SNe  show a large variety  of observed properties
\citep{turatto07},  depending  on   the  physical  parameters  of  the
progenitor  stars (radius,  mass, chemical  composition, circumstellar
environment)  at  the time  of  explosion.   The  variety of  stripped
envelope CC SNe became even more evident when in the late 1990's a few
SNe Ic with extremely broad spectral features were discovered, some of
which  had also  unusually  high luminosity  \citep[e.g.][]{galama98}.
This prompted the introduction of a new subclass, named broad-line SNe
Ic (SNe  BL-Ic).  The broad spectral  lines in SNe  BL-Ic suggest high
velocities  of the  expanding ejecta  and  a large  kinetic energy  to
ejected  mass ratio.   Recently  some objects  with yet  unprecedented
characteristics             have            been            discovered
\citep[e.g. SN\,2006jc,][]{pastorello07}.  In  this letter we focus on
the   early-phase  observations   of  SN\,2007gr,   another   type  Ic
stripped-envelope CC SN with interesting properties.

\section{Spectral evolution and line identification} \label{spectraevolution}

SN\,2007gr exploded in  NGC~1058 \citep{li07}, a member of  a group of
nearby  galaxies.   The distance  to  NGC  925,  one of  the  galaxies
belonging to the  group, was measured as 9.3\,Mpc  using Cepheid stars
\citep{silbermann96}.    We  adopted  this   distance  also   for  NGC
1058\footnote{A  different  distance (10.6$\pm$1.3  Mpc)  was used  by
\cite{crockett07} in a previous work  on SN 2007gr. With this distance
the  SN  would be  0.28  magnitude  brighter  than calculated  here.}.
SN\,2007gr is  thus one of  the nearest stripped-envelope CC  SNe ever
observed,    and   a   suitable    target   for    progenitor   search
\citep{crockett07}.  NGC~1058 is a prolific  host of SNe: two other CC
SNe (1961V and 1969L) occurred in the galaxy.

SN\,2007gr was discovered  on 2007 Aug 15.51 UT,  but was not detected
in  a  KAIT  image  taken   on  Aug  10.44  UT  \citep[unfiltered  mag
$<18.9$,][]{madison07}.  This constrains the  explosion epoch to $<$ 5
days  before discovery,  which by  itself  makes of  SN~2007gr a  very
interesting event.  Based on  a spectrum obtained the following night,
SN\,2007gr was classified as a generic  SN Ib/c since it was not clear
whether  the line  near  6350\AA\ was  really  \HeI\ (as  for SNe  Ib)
\citep{chornock07}. As  described below, the  later spectral evolution
did  not  confirm  the  presence  of  He. SN\,2007gr  is  thus  to  be
classified as type Ic.

Starting the night after  the discovery, we observed SN\,2007gr almost
every  night  until  it  reached  maximum brightness,  and  then  less
frequently  during the  decline phase.   Thanks to  the  good spectral
coverage during the rising phase,  unprecedented for a type Ic SN, the
origin of  the 6350\AA\ feature  can be investigated in  detail.  This
feature  is  visible  in  the  earliest spectra  of  several  SNe  Ic,
sometimes  partially blended  with a  feature near  6150\AA\  which is
often  attributed  to  \SiII   \footnote{Other  ions  have  been  also
suggested for  the $\sim  6150$\AA\ feature: \NeI\  \citep{sauer06} or
detached \Ha\ \citep{branch06}.}.

\begin{figure}
\includegraphics[width=8cm,height=8.5cm]{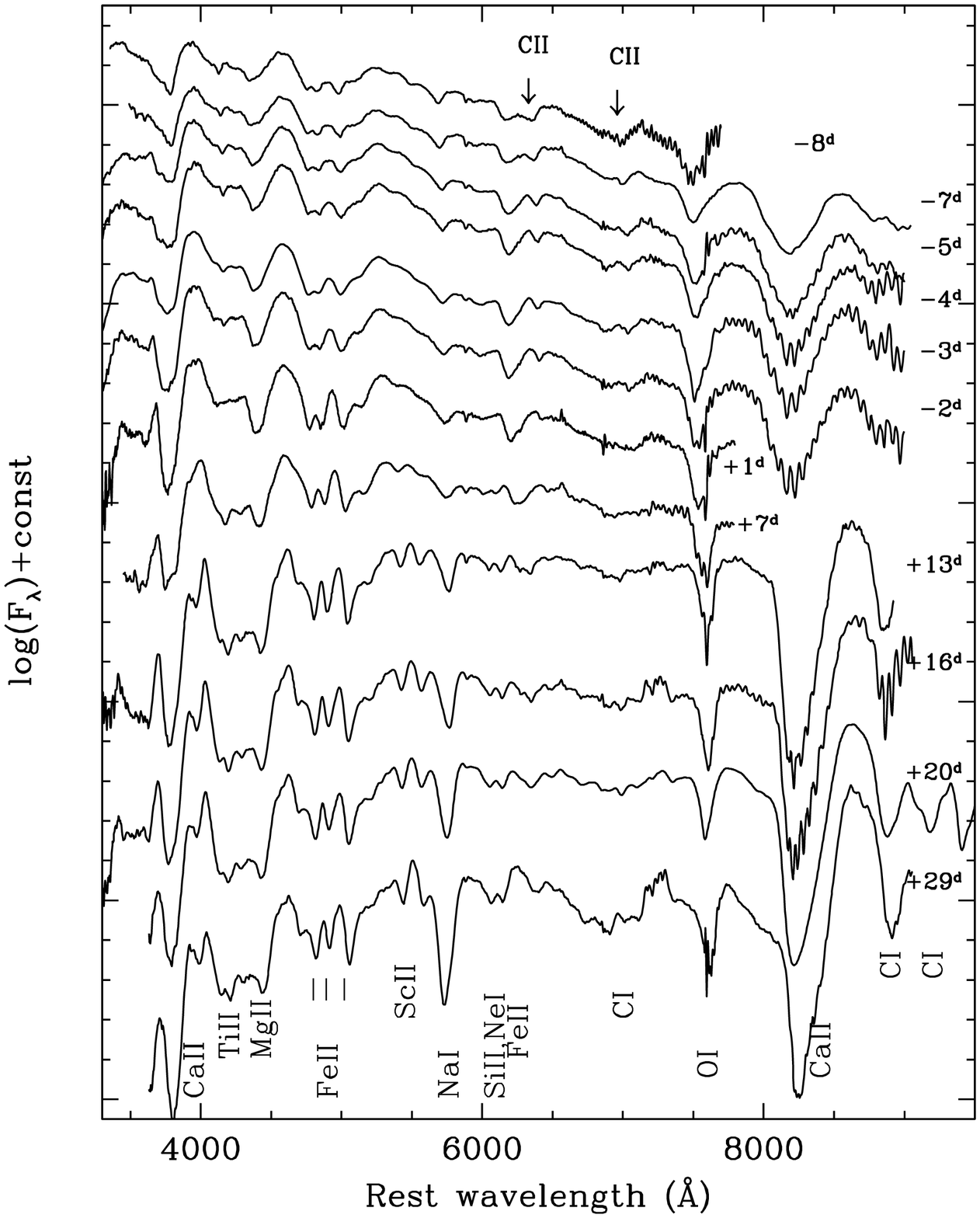}
\caption{Spectral evolution of SN\,2007gr.  The spectra are plotted in
the rest frame of the SN.} 
\label{fig2}
\end{figure}

Against  the   identification  of   the  6350\AA\  feature   as  \HeI\
\citep{chornock07} is the  fact that while in SNe  Ib the intensity of
the helium features  typically increases with time \citep{matheson01},
as  is expected because  of the  increased diffusion  of $\gamma$-rays
\citep{mazzali98a}, in  SN\,2007gr the feature decreases  and the line
even disappears  around maximum (see Fig. \ref{fig2}).   A more likely
identification  for   this  line  is  \CII\   $\lambda$6580  at  $\sim
11000$\,\kms.   This interpretation  is supported  by the  presence of
another line  at $\sim 7000$  \AA\ attributed to  \CII\ $\lambda$7235,
which  also  disappears around  maximum.   The  velocity  of \CII\  is
similar to  that of the  lines of other  ions.  The \CII\  features in
SN\,2007gr show an evolution similar to that seen in SNe Ia at similar
phases \citep{mazzali01}.

\begin{figure*}
\includegraphics[height=16cm, angle=-90]{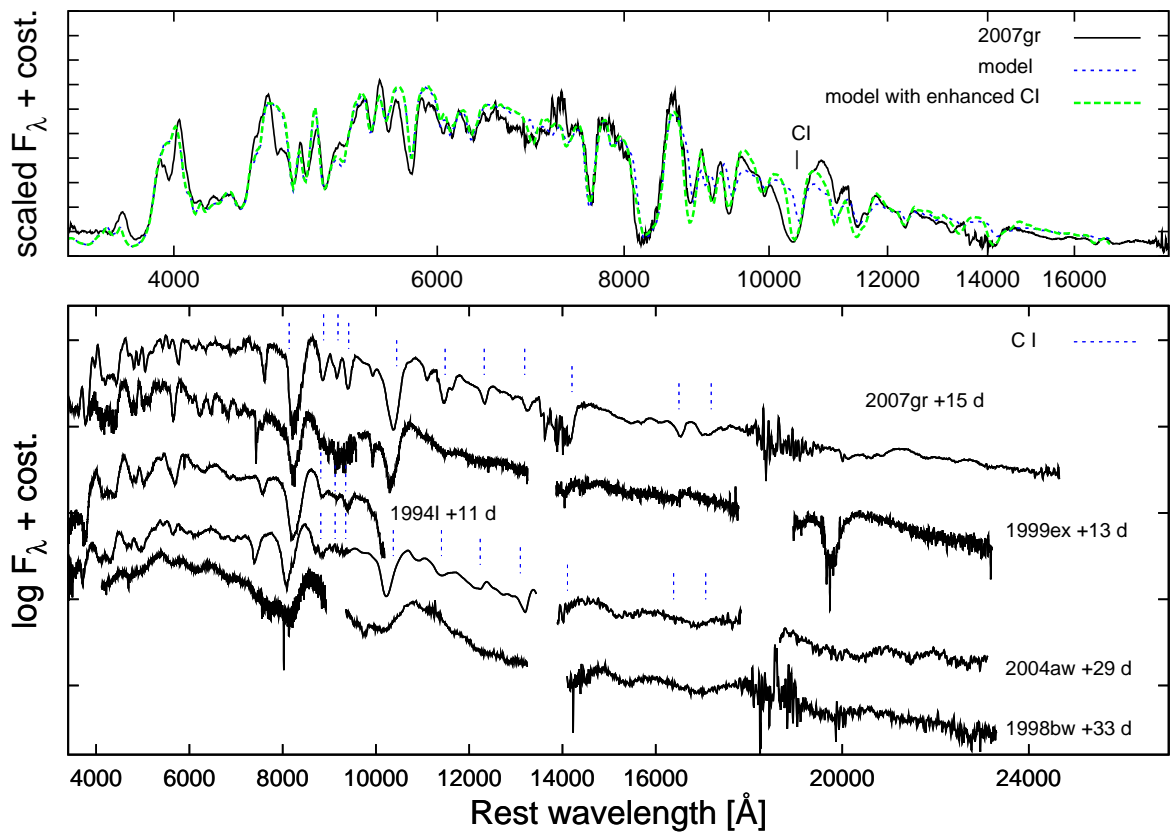}
\caption{Upper panel: spectrum of SN\,2007gr two weeks after B maximum
(solid  line) and  a synthetic  spectrum (dotted  line).   A synthetic
spectrum  with exnhanced  \CI\ is  also shown  (dashed line).   In the
lower  panel the  same  spectrum  is compared  with  spectra of  other
stripped-envelope CC  SNe. The positions of \CI\  features are marked.
SNe 2007gr, 1994I, and possibly 2004aw, show \CI\ features in the near
infrared, while SNe 1999ex and 1998bw do not.  However the presence of
\CI\  in  these  objects  can  not be  excluded.   In  particular,  in
SN\,1998bw the carbon features  could be present but strongly blended.
}
\label{fig4}
\end{figure*}

At later  epochs, after the  \CII\/ features fade, strong  \CI\/ lines
appear in the spectra, in  particular in the near infrared, indicating
a decrease of the temperature. This is shown in Fig.~\ref{fig4} (upper
panel) where  the optical and  infrared spectrum of SN\,2007gr  at two
weeks past maximum is compared with synthetic spectra calculated using
a        Monte       Carlo        spectrum        synthesis       code
\citep{mazzali93b,lucy99,mazzali00}.   We assumed a  power-law density
profile ($\rho\propto v^{-10}$)  and a lower boundary ``photospheric''
velocity  $v_{\rm ph}=6400\,${\kms},  with  an ejecta  mass above  the
photosphere of $1.4$ \Mo.

The  composition was  taken to  be  homogeneous for  all elements  and
includes $60\%$  O, $30\%$ C, $9.5\%$  Ne, a total of  $0.2\%$ Na, Mg,
Si, S, and Ca, and about  $0.28\%$ heavy elements (Sc, Ti, Cr, Fe, Co,
and Ni).

The model  (dotted line  in Fig.~2) provides  a reasonable fit  to the
observed spectrum. The absorptions seen in the IR are mostly caused by
\CI,  but the  observed  line  strength is  not  reproduced.  For  the
density structure used in the models the depth of the \CI\ features is
relatively  insensitive to  the  C mass  fraction  for mass  fractions
exceeding $\sim$$10\%$.  Therefore,  the simple density structure used
here may  need improvement, specifically  at high velocities  to allow
more carbon to recombine than predicted by the model.

To show  this more clearly, in  Fig. \ref{fig4}, we  also show (dashed
line)  a  model  where  the   optical  depth  in  all  \CI\  lines  is
artificially enhanced  at $v>8600\,$\kms  by factors of  20 to  100 at
different velocities.   The feature  at 10,400\,\AA, can  be explained
almost entirely  by \CI\  in this  way, except for  blue wing  that is
probably  due  to  other  ions  \citep[e.g.   \MgII,  \SiI\  or  \HeI,
][]{gerardy02,filippenko95}.

In  previous analyses  of SN~1994I,  the  origin of  the $10400$  \AA\
absorption was discussed, including  the possibility of a contribution
of  \CI\ \citep{millard99,sauer06}.  SN~2007gr  shows a  much stronger
signature   of  this  ion   than  other   events,  which   makes  this
identification more convincing.

A   common  alternative   identification   of  this   line  is   \HeI\
\citep{filippenko95}; however,  none of the other  \HeI\ features that
are  expected are  present  in  our observed  spectrum  to a  strength
sufficiently large to support  this identification. In particular, the
\HeI\ line  at 20580\AA\ should be  clearly visible if  \HeI\ were the
main contributor to the 10400\AA\ absorption.

Incidentally, among  the stripped-envelope CC SNe shown  in the bottom
panel of  Fig. \ref{fig4},  only SN\,1999ex (a  SN Ib) shows  a strong
\HeI\ $\lambda$20587.  The  other SNe are all of type  Ic, and none of
them show that feature.  The  \CI\ features are particularly strong in
SN\,2007gr, while they  are weaker in SNe 1994I  and 2004aw and absent
in SNe 1998bw, 2002ap, and  1999ex. SN\,1999ex may have burned most of
the  progenitor's carbon,  since at  the early  phases it  shows quite
strong  \MgII\ features \citep{hamuy02}.   Mg is  one of  the elements
produced in carbon burning \citep{woosley02}.

The low C abundance in the BL  SN Ic 2002ap is confirmed by a study of
the near-IR  spectra \citep{gerardy02}.  These authors  also noted the
presence of \CI\ in the infrared spectrum of the SN Ic 2000ew, and the
absence of strong \CI\/ in the SN Ib 2001B.

The emerging  evidence calls for a  complex scenario. It  is not clear
why some  SNe Ib  and BL SNe  Ic show  weak (if any)  carbon features,
while normal SNe  Ic show a wide range in C/O  optical lines.  The low
carbon abundance  in BL  SNe Ic  could be due  to more  massive and/or
lower  metallicity  progenitors  \citep{woosley02} than  for  standard
SNe~Ic.  However, it is difficult to place SNe Ib in the same context.

\section{Narrow lines and photospheric velocity}
\label{photoevol}

The identification  of \CI\ in  the infrared spectra of  SN\,2007gr is
facilitated  by the fact  that the  spectral features  in this  SN are
narrower than in  other SNe Ic (for instance,  the \FeII\/ features at
4924, 5018, 5170\AA\ are unblended, which is very uncommon for type Ic
SNe).

The  narrow  lines  are  probably  related to  the  particularly  slow
evolution of  the photospheric expansion velocity  of SN\,2007gr which
ranges from  11000\,\kms\ one week  after explosion to  4800\,\kms\ at
$\sim  50$ days  after explosion.   The small  range of  velocities at
which the  lines could  form makes them  narrow compared to  other SNe
Ic.  By  contrast,  SN\,2002ap  reaches  even  lower  velocities  than
SN\,2007gr at late phases, but it still shows broad features.  This is
due to  the shallow  density gradient and  the huge velocity  range at
which the lines can form in  SN\,2002ap.  In fact, at early phases the
velocity derived for SN\,2002ap is $\sim 5000$\,\kms\ higher than that
obtained   for  SN\,2007gr.    No   data  are   available  for   other
stripped-envelope CC SNe Ic at such an early epoch.

The fact  that at $\sim 50$  days after explosion  SN\,2007gr is still
optically  thick  inside 5000\,\kms\  and  that  there  is not  enough
material  outside   11000\,\kms\  to  form  lines   may  suggest  that
SN\,2007gr is a compact object  inside $\sim 5000$\,\kms, with a sharp
tail out to 11000\,\kms.

\section{Light curve}
\label{lightcurve}

SN\,2007gr  reached  maximum  on  Aug  28  at  $m_\mathrm{R}=  12.77$.
Adopting  a  distance  modulus  of  29.84, a  galactic  extinction  of
E(B-V)=0.062  \citep{schlegel98},  and  a  host galaxy  extinction  of
E(B-V)=0.03, the absolute R  magnitude is -17.3 (see Fig. \ref{fig1}).
The  estimate  of the  host  galaxy  extinction  is derived  from  the
relative intensity of the  galactic and host galaxy interstellar \NaID
1 absorption  lines, assuming that  the galactic and host  galaxy dust
properties are similar.  Equivalent widths of 0.31 and 0.13\,\AA\ were
measured  for  the  galactic  and  host galaxy  components  of  \NaID,
respectively,  in a  high-resolution spectrum  obtained with  the FIES
Echelle spectrograph at NOT.

The  rise time  of SN\,2007gr  is  constrained by  the KAIT  telescope
non-detection of 10 August.  Thus the rise time to R maximum turns out
to be  14-18 days. The  rise time to  B maximum (which occurred  on 24
August)  is  10-14  days,  intermediate between  those  of  SN\,2002ap
\citep[8      days,][]{foley03}      and     SN\,1999ex      \citep[18
days,][]{stritzinger02}.  After  maximum, the luminosity  decline rate
is similar  to that  of SNe  2002ap and 1999ex,  while SNe  2004aw and
1998bw are slower, and SN\,1994I is faster.

\section{Discussion}

From  the  light  curve   comparison  and  the  photospheric  velocity
evolution, we  can get some  estimates for the physical  parameters of
SN\,2007gr.

The absolute  R maximum of SN\,2007gr,  $-17.3$ is similar  to that of
SN\,2002ap ($M_{R}  = -17.4$).  Therefore these SNe  likely produced a
similar amount of $^{56}$Ni ($\sim$ 0.07-0.1 \Mo), SN\,2007gr possibly
slightly more than SN\,2002ap owing to the longer rise time.

The width of the light curve $\tau_\mathrm{peak}$ and the photospheric
velocity  $v_\mathrm{ph}$ around  maximum depend  on the  ejected mass
$M_\mathrm{ej}$     and    the    kinetic     energy    $E_\mathrm{k}$
\citep{arnett82,arnett96} as follows:

$\tau_{peak}~\propto~M^{3/4}_{ej}  E^{-1/4}_{k}$  and $v_{ph}~\propto~
M_{ej}^{-1/2} E_{k}^{+1/2}$.

Since  the  photospheric  velocity  of SN\,2007gr  around  maximum  is
similar to that of SN~1994I, while  the peak of the light curve is 1.5
times broader, we expect an ejected  mass of $\sim 1.5-3$ \Mo\/, and a
kinetic  energy $\sim 1.5-3  \times 10^{51}$  erg\footnote{The ejected
mass and  kinetic energy  are derived scaling  the values  obtained by
\cite{sauer06}  and   \cite{mazzali07}  for  SNe   1994I  and  2002ap,
respectively.}.

A more robust estimate of the  ejected mass and kinetic energy will be
derived   from  the   nebular  data   and  their   detailed  modelling
\citep[e.g.][]{mazzali07},  but   our  preliminary  estimates  already
suggest that SN\,2007gr resembles  SN\,2002ap in terms of mass ejected
even  though the  kinetic  energy is  a  factor of  two smaller.  This
combination makes SN\,2007gr similar  to SN\,1994I with respect to the
ratio of  kinetic energy to ejected  mass, but with  a steeper density
profile, resulting in narrow lines.

The analysis of the photospheric phase of SN\,2007gr confirms the wide
variety of  Ic SNe,  not only  in light-curve shape  and width  of the
spectral features, but also in  the range of carbon abundances and the
shape of underlying density profiles.

If asymmetry  is the  key to explain  this heterogeneity,  the similar
light curves of SN\,2007gr  and SN\,2002ap combined with the different
density profiles (compact for  SN\,2007gr and extended for SN\,2002ap)
may  suggest that  SN\,2007gr and  SN\,2002ap were  asymmetric bipolar
explosions observed  close to the minor and  major axis, respectively.
Nebular spectra may provide  important information to corroborate this
hypothesis.    Other  scenarios  to   explain  the   heterogeneity  of
stripped-envelope CC SNe (different progenitor, radius, mass, chemical
composition,  circumstellar  environment) cannot  be  excluded on  the
basis of the present data.

\acknowledgments  We thank an anonymous referee and
 L.   Occhionero, R.   Bogh,  C.  Gall,  G.
Leloudas, M.   Ladegaard, P. Krogstrup, T. Hansen,  J.G.M. Samsing, D.
Juncher, K.A. Gregersen, K.P.  Olsen, L.A. Thomsen, L.O.  S{\o}rensen,
J. Ohrt, T.A.  Ottosen and  J.C.H. Riggelsen for observing 07gr at NOT
during  two Summer Schools.  S.V.  thank  F.  Frontera  for continuous
support.   This  work  was  supported  by grants  n.   2005025417  and
2006022731  of the  PRIN of  Italian Ministry  of University  and Sci.
Reaserch.   The work  of  D.T.   was partly  supported  by RFBR  grant
05-02-17480. The  Dark Cosmology Center  is funded by  Danish National
Research Foundation.  This paper is based on  observations made during
the following program: AOT16\-TAC\_52 (TNG), Benetti et al (Ekar).

{\it    Facilities:}   \facility{Asiago    Ekar  Telescope},
\facility{Asiago   Pennar   Telescope},   \facility{Telescopio
Nazionale     Galileo},     \facility{Nordic    Optical     Telescope},
\facility{Wendelstein Telescope}, \facility{Sternberg Astronomical
Institute Telescope}.

\clearpage

\begin{figure}
\includegraphics[width=5.5cm,height=8cm, angle=-90]{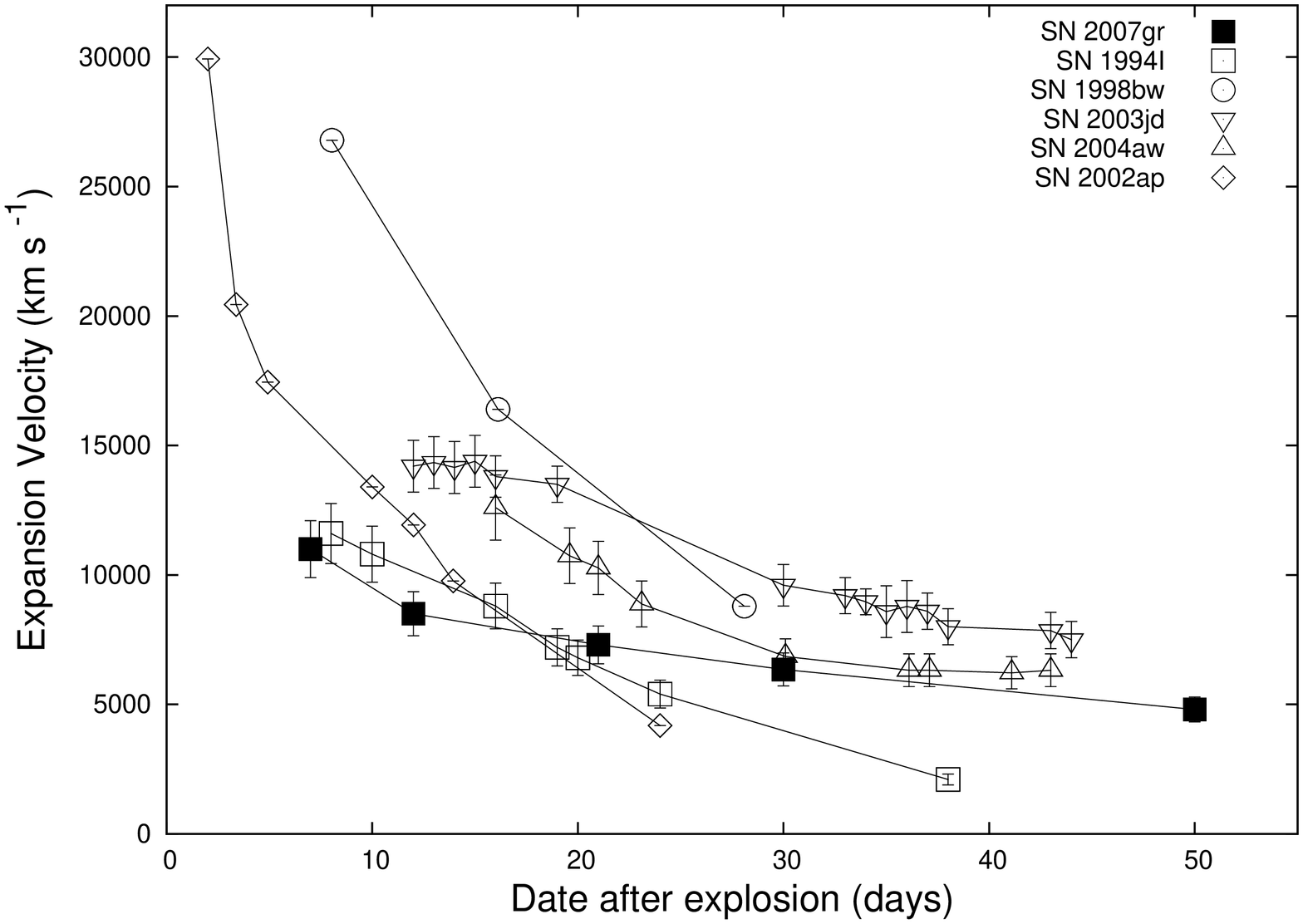}
\caption{The photospheric velocity of SN\,2007gr and of other stripped
envelope CC SNe are  shown.  The photospheric velocities of SN\,2007gr
is from preliminar spectral modelling.  Those of SNe 1994I, 1998bw and
2002ap    are   from    spectral    modelling   by    \citet{sauer06},
\citet{iwamoto98} and \citet{mazzali02}.  The velocities of SNe 2003jd
and  2004aw  are  those  where  the \SiII\  $\lambda$6355  lines  form
\citep{valenti07,taubenberger06},  which  is  a  good  tracer  of  the
photospheric velocity.}
\label{fig3}
\end{figure}

\begin{figure}
\includegraphics[width=5.cm,height=8cm, angle=-90]{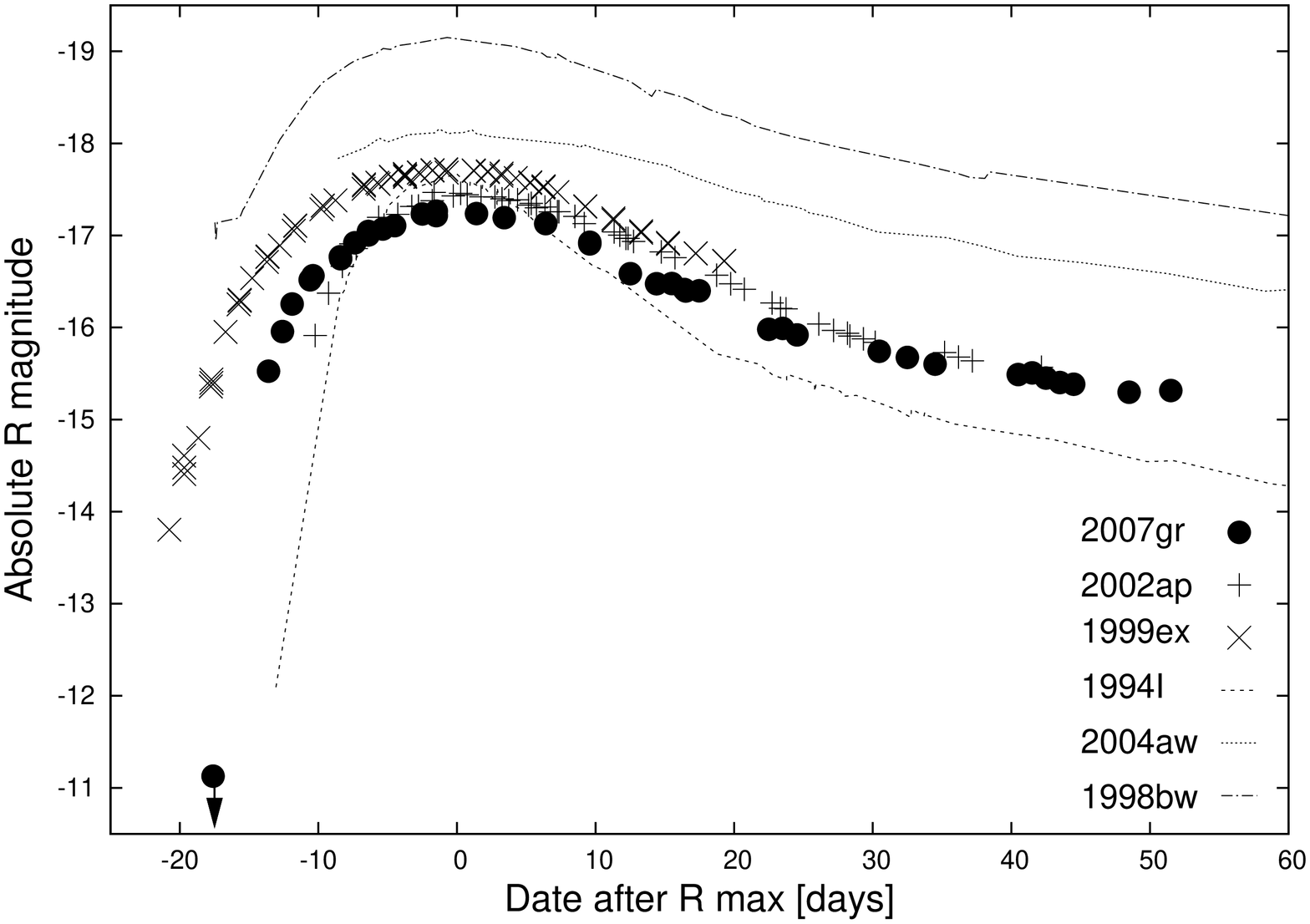}
\caption{The R  light curve of  SN\,2007gr is shown together  with the
light  curves  of other  stripped-envelope  CC  SNe.  Color  excesses,
distance moduli  and data  references for the  SNe are as  follows: SN
2007gr: E(B-V)=0.09,  $\mu$=29.8 this letter;  SN 2003jd: E(B-V)=0.14,
$\mu$=34.46,  \citet{valenti07}; SN 1998bw:  E(B-V)=0.06, $\mu$=32.76,
\citet{galama98,patat01};   SN    2002ap:   E(B-V)=0.09,   $\mu$=29.5,
\citet{yoshii03,foley03};   SN   1999ex:   E(B-V)=0.30,   $\mu$=33.28,
\citet{stritzinger02};     SN    1994I:     E(B-V)=0.3,    $\mu$=29.6,
\citet{richmond96};     SN     2004aw    E(B-V)=0.37,     $\mu$=34.17,
\citet{taubenberger06}.}
\label{fig1}
\end{figure}

\end{document}